%% file: article.tex
\documentclass[british,twocolumn]{ECMS}
\usepackage{multicol}
\usepackage{rotating}
\usepackage{lmodern}
\usepackage{babel}
\usepackage[utf8]{inputenc}
\usepackage[T1]{fontenc}
\usepackage{url}
\usepackage{csquotes}

\newtheorem{remark}{Remark}

\usepackage{comment}
\usepackage{graphicx}
\usepackage{amsmath,amssymb}
\usepackage{subcaption}

\usepackage{url}
\usepackage[vskip=0pt,font=itshape]{quoting}

\usepackage[backend=biber, natbib=true, style=apa]{biblatex}

\title{On the Simulation of Gravitational Lensing}
\author{
   \begin{tabular}{cccccc}
	   Hans Georg Schaathun$^1$ &
	   Ben David Normann &
	   Einar Leite Austnes \\
     $\langle$georg@schaathun.net$\rangle$ &
     $\langle$ben.d.normann@ntnu.no$\rangle$ &
     $\langle$eaustnes@gmail.com$\rangle$ 
           \\[7pt]
	   Simon Ingebrigtsen &
           Sondre Westbø Remøy &
	   Simon Nedreberg Runde \\
     $\langle$simon.ing.89@gmail.com$\rangle$ &
           $\langle$sore@live.no$\rangle$  &
     $\langle$simonrun@hotmail.com$\rangle$ \\
   \end{tabular}
   \\
   Department of ICT and Natural Sciences \\
   Faculty of Information Technology and Electrical Engineering \\
   N-6025 Ålesund, Norway\\
   \footnotesize
   $^1$Corresponding author
   }
\begin{document}
\maketitle
\thispagestyle{empty}\pagestyle{empty}

\input main

%\footnotesize
\printbibliography

\end{document}

%% file: main.tex
% \begin{abstract}
% \end{abstract}

\section*{Keywords}

Gravitational Lensing; Simulation; Dark Matter

\section*{Abstract}

Gravitational lensing refers to the deflection of light by the
gravity of celestial bodies, often predominantly composed of
dark matter.  Seen through a gravitational lens, the images
of distant galaxies appear distorted.  
In this paper we discuss simulation of the image distortion
by gravitational lensing.
The objective is to enhance our understanding of how gravitational
lensing works through a simple tool to visualise hypotheses.
The simulator can also generate synthetic data for the purpose
of machine learning, which will hopefully allow us to invert
the distortion function, something which is not analytically
possible at present.

\section{Introduction}

One of the big questions in astrophysics is the mapping of the Universe.
Modern telescopes provide enormous amounts of images of the night sky,
but about $85\%$ of the mass is dark matter (DM).  Emitting no light,
this dark matter is not visible on the images.  However, because of
gravity, the dark matter can distort the light from more remote objects.
This is called a gravitational lens (GL)~\citep[e.g.][]{bertone18}, because
it works analogously to an optical lens.

The shape and location of gravitational lenses can be inferred by studying
images of galaxies which appear distorted from our viewpoint,
but the calculations are complex and may amount to days of
manual work for a single lens.
Attempts to automate this process, for instance using machine
learning \citep[e.g.][]{hezaveh17}, are promising but still
limited to selected cases.

In this paper we develop a framework for simulating gravitational lenses,
that is, to synthesise authentic distorted images, given an undistorted
source image and a lens model.
This has two purposes.  Firstly, it enables bulk generation of 
synthetic data sets which can be used for training in machine learning.
Secondly, it provides a graphical user interface where cosmologists can
experiment to test and explore hypotheses.
As simulation model we use the Roulettes formalism \citep{clarkson16a},
which is notable by unifying weak and strong lensing in one paradigm.
This formalism provides the forward calculation of the distorted image,
but is not analytically invertible, which means that we cannot immediately
infer the lens or source profiles from the distorted image.

\input bendavid

\input model

\section{The simulator software}

The simulator works with pixmap representations of the 
source image and the distorted image.
The Roulettes model maps Cartesian co-ordinates in the 
lens plane to polar co-ordinates in the source plane.
Hence it is trivial to generate the distorted image pixel
by pixel, by simply looking up the corresponding pixel
(light ray) in the source image.
Fractional pixel co-ordinates may be interpolated, but if high-resolution
images are used, this is not necessary.
Even though the distorted image is calculated in the lens plane 
according to the Roulettes formalism,
we project it back into the source plane,
so that the scale (size) is comparable to the source image.

\begin{figure*}
   \begin{center}
      \begin{minipage}{0.8\textwidth}
         \includegraphics[width=\textwidth]{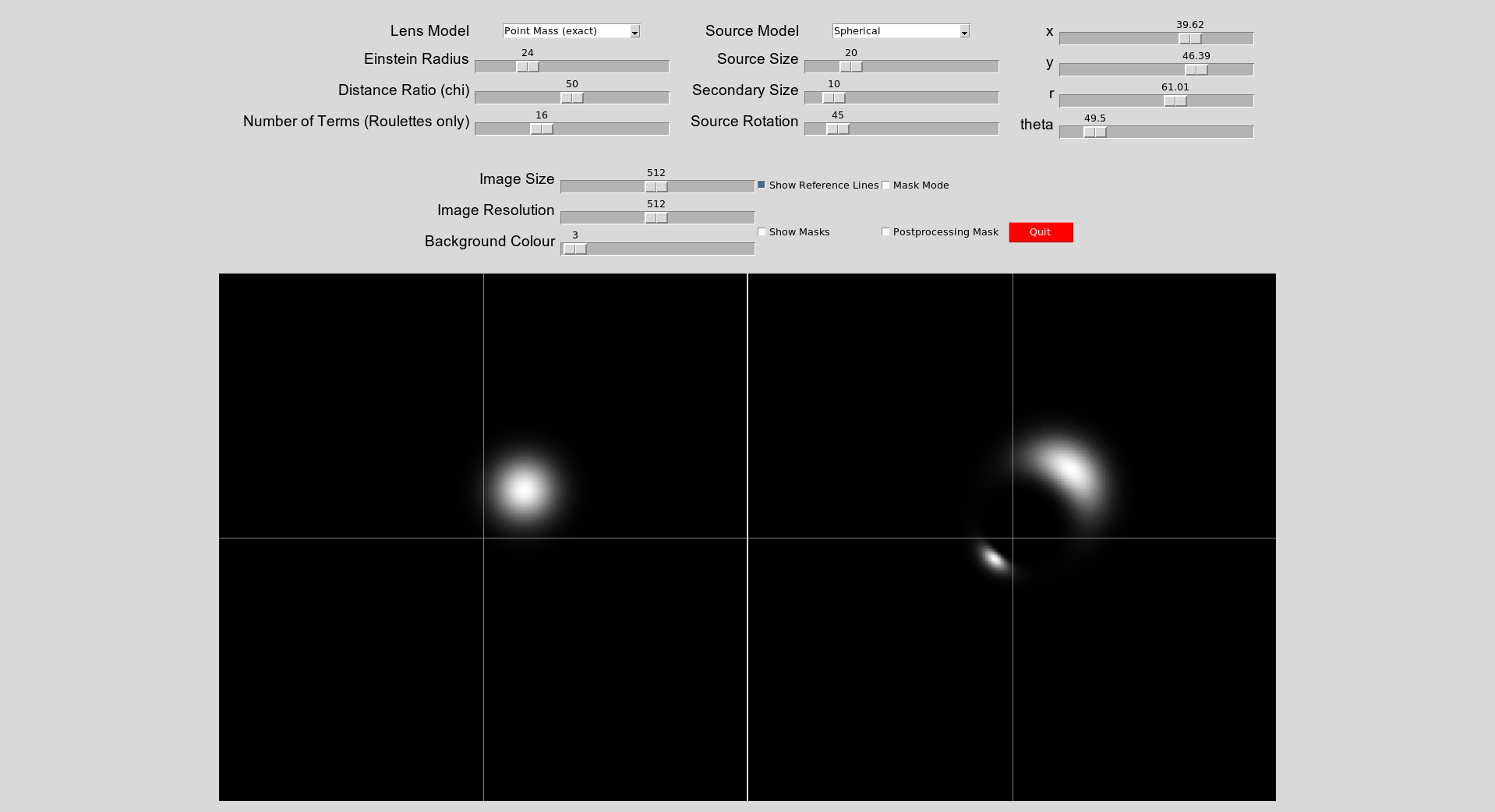}
      \end{minipage}
   \end{center}
   \caption{The GUI for the Simulator.}
   \label{fig:gui}
\end{figure*}

The simulator is implemented as a C++ library, using OpenCV for
image manipulation.  Front-end tools are implemted in Python, 
using Pybind11 to wrap the C++ library.
There is a GUI tool, as shown in Figure~\ref{fig:gui}, and a command
line tool to generate images in bulk.
The software is available in Open Source on github\footnote{%
   The release used in this paper is v2.0.2 at
   \url{https://github.com/CosmoAI-AES/CosmoSim/releases/tag/v2.0.2}.}.

% \subsection{Simulator Architecture}

The simulator is a very simple object-oriented structure, where new
lens and source models can easily be added.
The abstract \emph{Source} class represents the source image,
with concrete subclasses for spherical and ellipsoid lenses.
These classes store the source image which is generated upon
instantiation.
The abstract \emph{LensModel} class represents the gravitational
lens with subclasses for point mass and spherical (SIS) lenses.
These classes implement the distortion function 
${\mathcal{D}}(r,\phi)$,
and store the distorted image as well as a reference to the source object, 
An update method computes the distorted image, which can be retrieved
with a getter function.

The python wrapper does not expose the object model.  The \emph{CosmoSim}
class has setters for types of lens and source models as well as all the 
relevant parameters.
It exposes the Lens Model's update method and getters for
the distorted and the actual image.
This reduces code size and simplifies maintenance, since we do not have
to keep wrapper classes for all the classes in the C++ library.
Still it gives complete access to all the features of the simulator.

A critical step in the SIS model is to calculate all the
amplitudes $\alpha_s^m$ and $\beta_s^m$.
We use Python to pre-generate expressions for each $(m,s)$
pair up to some maximum truncation limit $m_0$, using the sympy
module to differentiate $\psi$.
The resulting algebraic expressions are loaded by the C++
code from a text file and evaluated numerically using the
symengine library.

\section{Results}

\begin{figure}
   \begin{subfigure}{41mm}
      \includegraphics[height=40mm]{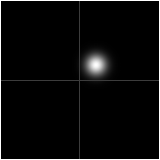}
      \caption{Source Image}
      \label{fig1pm}
   \end{subfigure}
      \hfill
   \begin{subfigure}{41mm}
      \includegraphics[height=40mm]{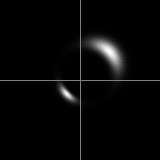}
      \caption{Exact model}
      \label{fig2pm}
   \end{subfigure}
   \begin{subfigure}{41mm}
      \includegraphics[height=40mm]{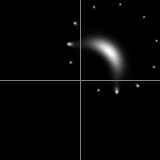}
      \caption{Roulettes with 10 terms}
      \label{fig3pm}
   \end{subfigure}
      \hfill
   \begin{subfigure}{41mm}
      \includegraphics[height=40mm]{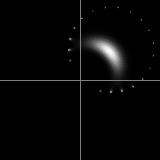}
      \caption{Roulettes with 20 terms}
      \label{fig4pm}
   \end{subfigure}
   \caption{Examples with a spherical source and point mass lens;
      $D_\textrm{L}/D_\textrm{S}=50\%$, $\xi=22$, $\theta=45^\circ$, $R_\textrm{E}=14$, $\sigma=7$.}
   \label{fig:pm}
\end{figure}

The GUI interface (Figure~\ref{fig:gui}) allows the user quickly 
to experiment with different parameter settings, and visually
review resulting distorted images.
For the cosmologist on the team, this has proved an invaluable tool,
particularly to develop intuition and develop a deeper understanding
of both the phenomenon (GL) and the model (Roulettes).
A particular point where it proved useful was in understanding the
convergence ring and the spurious images which we discuss below.
Moreover, it has allowed us to verify the theory.

The spurious images is a model artifact.
Calculating the distorted images in the Roulettes formalism with
an even truncation threshold $m_0$ produces $m_0+1$ spurious
images in a ring roughly centred on the local origin $\boldsymbol{\xi}$.
The model is exact at the origin, and a good approximation in a 
neighhood around it.
This is clearly seen in the comparison of the exact point mass model
and the Roulettes approximations in Figure~\ref{fig:pm}.
On one hand, these simulations show how well the Roulettes formalism
matches the exact solution, something which can also be verified
quantitatively by computing difference images.
On the other hand it illustrates the convergence ring, outside of which
the model is meaningless, with the spurious images as a blatant example.

Asymptotically, when the number of terms $m_0$ tends to infinity, it can be
shown that this ring has radius $\xi$ centred on $\boldsymbol{\xi}$,
and that it approaches the limit from the outside.
This result is provided by~\citet{Clarkson_2016_II} 
and is called the ring of convergence.
We can also see in Figure~\ref{fig:pm} how the spurious images are
smaller for large $m_0$, as the light is distributed between more images.
When the number of images tends to infinity, the size of each one will
tend to zero.

\begin{figure}
   \begin{subfigure}{41mm}
      \includegraphics[height=40mm]{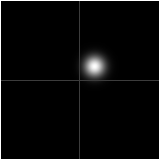}
      \caption{Source Image $\xi=20$}
      \label{fig1sis}
   \end{subfigure}
      \hfill
   \begin{subfigure}{41mm}
      \includegraphics[height=40mm]{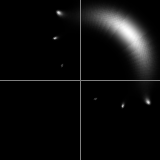}
      \caption{Distorted image $\xi=20$}
      \label{fig2sis}
   \end{subfigure}
   \begin{subfigure}{41mm}
      \includegraphics[height=40mm]{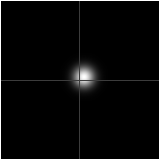}
      \caption{Source Image $\xi=5$}
      \label{fig3sis}
   \end{subfigure}
      \hfill
   \begin{subfigure}{41mm}
      \includegraphics[height=40mm]{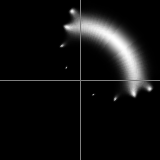}
      \caption{Distorted image $\xi=5$}
      \label{fig4sis}
   \end{subfigure}
   \caption{Examples with a spherical source and SIS lens,
      with different source positions;
      $D_\textrm{L}/D_\textrm{S}=50\%$, $\theta=45^\circ$, $\xi=24$, $\sigma=7$.}
   \label{fig:sis}
\end{figure}

Figure~\ref{fig:sis} shows an example of the behaviour for
different degrees on lensing.
When the distance $\xi$ between the lens and the distorted 
images is smaller, compared to the Einstein radius $R_\textrm{E}$, the
lensing effect is weaker.  If it is sufficiently small, the image
fits well inside the convergence ring and is a good representation
of the physical behaviour.
For stronger lensing effects (Figure~\ref{fig4sis}), we can see how
the image is drawn out towards the spurious image.
Thus we have demonstrated a limit for when the Roulettes formalism 
is satisfactory.
Even though the Roulettes unify weak and strong lensing in one
paradigm, it is not yet satisfactory in the case of very strong lenses.

\begin{figure}
   \begin{subfigure}{41mm}
      \includegraphics[height=40mm]{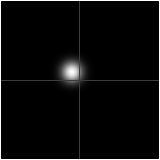}
      \caption{Source Image}
      \label{fig1ring}
   \end{subfigure}
      \hfill
   \begin{subfigure}{41mm}
      \includegraphics[height=40mm]{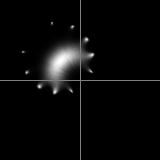}
      \caption{10 terms}
      \label{fig2ring}
   \end{subfigure}
   \begin{subfigure}{41mm}
      \includegraphics[height=40mm]{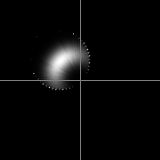}
      \caption{50 terms}
      \label{fig3ring}
   \end{subfigure}
      \hfill
   \begin{subfigure}{41mm}
      \includegraphics[height=40mm]{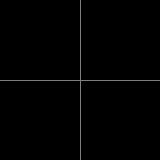}
      \caption{150 terms}
      \label{fig4ring}
   \end{subfigure}
   \caption{Examples of the spurious images for various numbers 
      of terms;
      $D_\textrm{L}/D_\textrm{S}=50\%$, $\theta=135^\circ$, $\xi=12$, $R_\textrm{E}=8$, $\sigma=7$.}
   \label{fig:ring}
\end{figure}

Knowing the shape of the convergence ring, it is possible to mask away
everything close to or outside the ring.
This is important to speed up the simulation, as the distortion equation
\eqref{eqn:general mapping} is computationally expensive, although
it depends heavily on the image size and on $m_0$.
However, there is no reason to calculate pixels outside the convergence
ring, and taking this into account, we get a reasonably responsive GUI
for image size $512\times512$ and $m_0=50$.
This masking is made optional in the tools.
Without the masking, the GUI is usuable around $m_0=16$, but it quickly
gets irresponsive for $m_0\ge20$.
This tests have used a desktop computer with
an AMD Ryzen 9 5900X 12-Core Processor at 2195.8MHz.
The image size of $512\times 512$ is, of course, a lot higher 
than typical empirical images, but the high resolution may be 
important for the testing of the theory.

For a more objective performance test, we have done bulk
generation of images, using the same desktop computer.
Generating 1000 images at $400\times400$ resolution took
35½s walltime and 10 minutes 5 seconds CPU time for $m_0=16$.
% 32s walltime and 9 minutes 4 seconds CPU time for $m_0=15$.
For $m_0=50$, it took
4 minutes 38 seconds walltime and 81 minutes 24 seconds CPU time,
and for $m_0=150$, $44'41''$ waltime and 11h19 CPU time.
This is very acceptable, although interactive applications may
not be able to go much above $m_0=50$.
For the purpose of machine learning, the training set 
generation is negligible compared to the training time, 
as it should be.

\section{Impact and Conclusion}

Our simulation model provides a computational representation of the
algebraic Roulettes formalism~\citep{clarkson16a}.
An important motivation has been to bridge the gap between computer
scientists and physicists, by developing a model which is meaningful 
in both domains.
This is a necessary first step to open up this important research
field from cosmology for a wider community, most importantly for
machine learning which may be able to invert the distortion function.

Similar simulators have been reported in the literature, each with their
own limitations.
The lenstronomy package \citep{birrer18} is a comprehensive package,
with other features in addition to the generation of distorted images,
but it is limited to strong lensing.
Other works we have found do not provide source code or sufficient detail
to reproduce it, making them difficult to validate or extend.
Thus the present transparent simulation model with suppor for both weak
and strong lensing is a considerable step forward.
We provide the first computational implementation the Roulettes formalism, 
and also a framework which can be extended with new lens modelles,
be they expressed in the Roulettes formalism or other frameworks.

We have not given any results on machine learning.  The first rudimentary
tests are promising, but more work is needed before it is ready for discussion. 
The simulator has other uses, as a visual tool for testing and exploring
hypotheses in cosmology.  Somewhat unanticipated, our simulations have 
revealed problems and limitations in the Roulettes formalism, and thus
identified needs for further research.

This work is a mere starting point, leaving several interesting open problems.
Development of machine learning models to reconstruct the lens profile and 
possibly the source image has already been mentioned.
To achieve this, we will also have to adapt our system to simulate the noisy,
low-resolution data in real images of the night sky.
An independent line of research is computational models for a broader range of
lens models.
It would be particularly interesting if we could use sampled representations
of the lens potential $\psi$.

\section*{Acknowledgements}

The first prototype of the simulator was a final year project (BSc)
by four of the authors \citep{cosmoai2022bsc}, and the work would have 
been impossible without their initial work.
The models and prototypes have since been extended and improved
by the other two authors. Also, we would like to thank Chris Clarkson at QMUL for his various inputs.

%% file: bendavid.tex
\section{Background on Gravitational Lensing}

All matter, ordinary or dark alike, acts as a lens, distorting the images of distant galaxies. In 1919, the deflection of light by the sun was measured by Eddington during a solar eclipse, and shown to agree with Einstein's theory of general relativity. Since then, theoretical work on lensing has been extensively developed. The scarcity and low resolution of observations, have however caused pessimism concerning the applicability of this tool for actual observation.
But one step at a time, the cosmological community has found ways to observe the phenomenon,
and at present, it is booming both with applications and observation.
Indeed, GL has become one of the major tools for mining information from the night sky.

One of the clear applications of GL is in understanding the nature of dark matter, which according to the present paradigm of cosmology is one of the main constituents in the universe. The other two are ordinary luminous matter ($\sim\,5\%$) and the so-called dark energy ($\sim 68\%$). While the former is the stuff that make up stars, planets and all the rest, dark energy is what causes the accelerated expansion of the late universe. Finally, dark matter ($\sim 27\%$) is the name given to mass indirectly observed in galaxies, but yet not seen. Its elusive nature has haunted cosmology since the 1930s.
Although dark matter does not emit light, it must have mass, and thus it bends light like ordinary (so-called luminous) matter. This means that a study of lensing by a distant galaxy is implicitly a study of the dark matter in the lens. By studying lenses at different locations in the sky, one may thus create maps of the distribution of dark matter in the universe. This is important in order to understand the nature of dark matter, since the distribution of a substance says much about the nature of it.

Traditionally, the algebraic framework for GL is divided into two regimes;
\textit{weak} and \textit{strong} GL, depending on the level of distortion.
These are two sides of the same physical phenomenon, and both types may be used to reconstruct DM distribution in cosmological surveys. It is therefore unfortunate, and artificial, that both lensing effects are not studied within the same algebraic framework. Especially when it comes to cluster-lenses, where effects from both regimes appear.

\citet{clarkson16a,Clarkson_2016_II}
found a way of solving this problem by developing an algebraic framework that can model both weak and strong GL,
henceforth referred to as the  \emph{Roulette formalism}. 
In theory, it should be possible to use the Roulette formalism to reconstruct the lens mass from images of distant galaxies subject to GL.
The usefulness of this formalism is also seen in its ability to go beyond shear 
measurements, by incorporating higher-order effects (to arbitrary order).
The ability to do so could prove useful as data received from the night 
sky drastically increases in amount and accuracy.

In order to make use of the Roulette formalism to such ends,
a number of questions should be answered. 
Firstly, since the Roulette formalism builds on a series expansion
which has to be truncated, it is important to know if the region
of convergence is large enough to give satisfactory images in 
practice.
Secondly, is it possible to generate images at a reasonable speed? 
In this work we answer these questions by implementing numerical computation
of the Roulettes formalism.

%% file: model.tex
\section{The Roulette formalism and its computation}

\begin{figure}%[t]
   \begin{center}
\includegraphics[width=0.9\columnwidth]{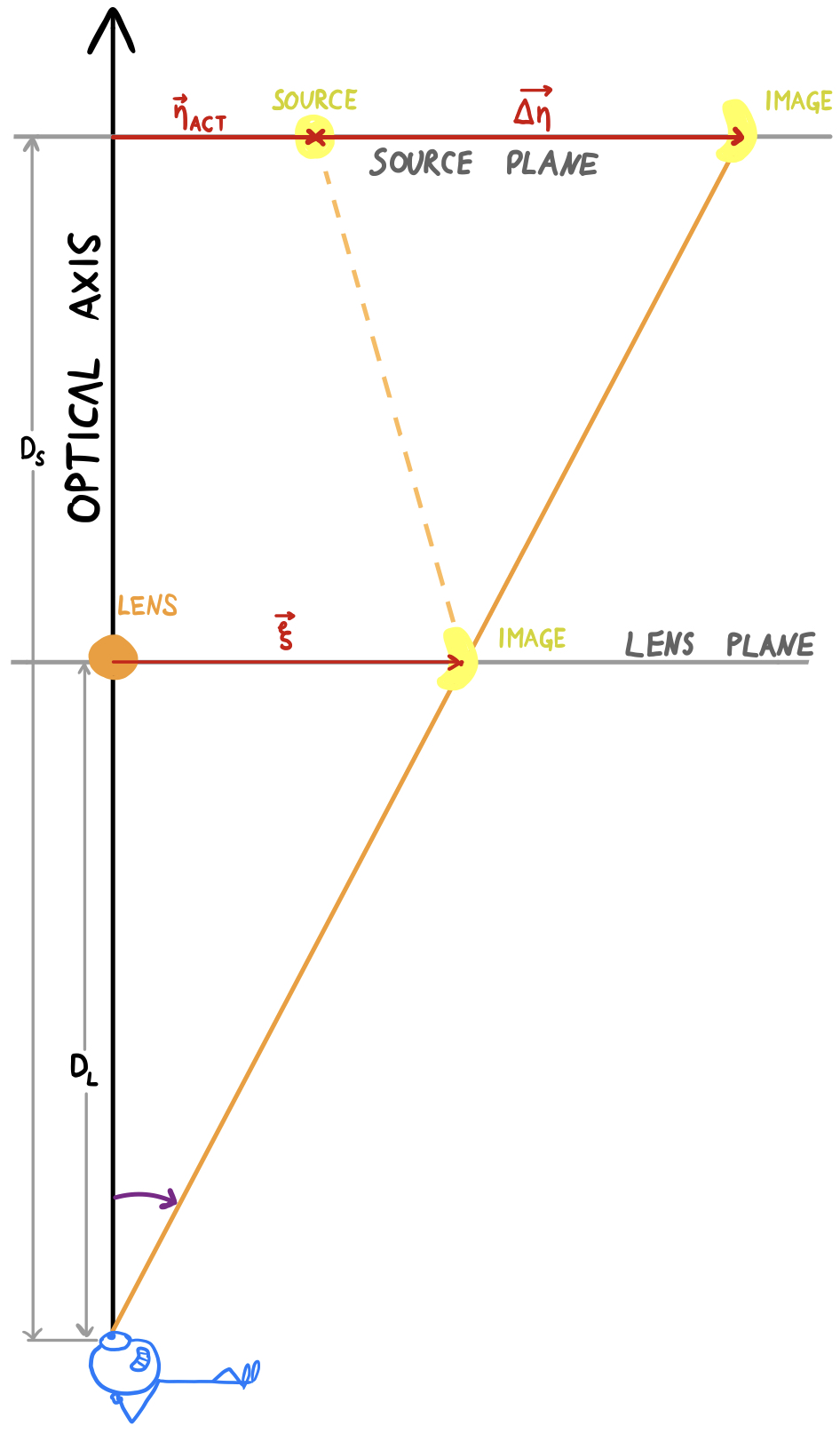}
   \end{center}
    \caption{The figure shows the set-up for the flat-sky approximation, with the source plane (the lens plane) a distance $D_{\mathrm{S}}$ ($D_{\mathrm{L}}$) from the observer. Compare with Figure~\ref{fig1} for more details.}
 \label{fig:model_2D}
\end{figure}
   
\begin{figure*}%[t]
   \begin{center}
\includegraphics[width=0.9\textwidth]{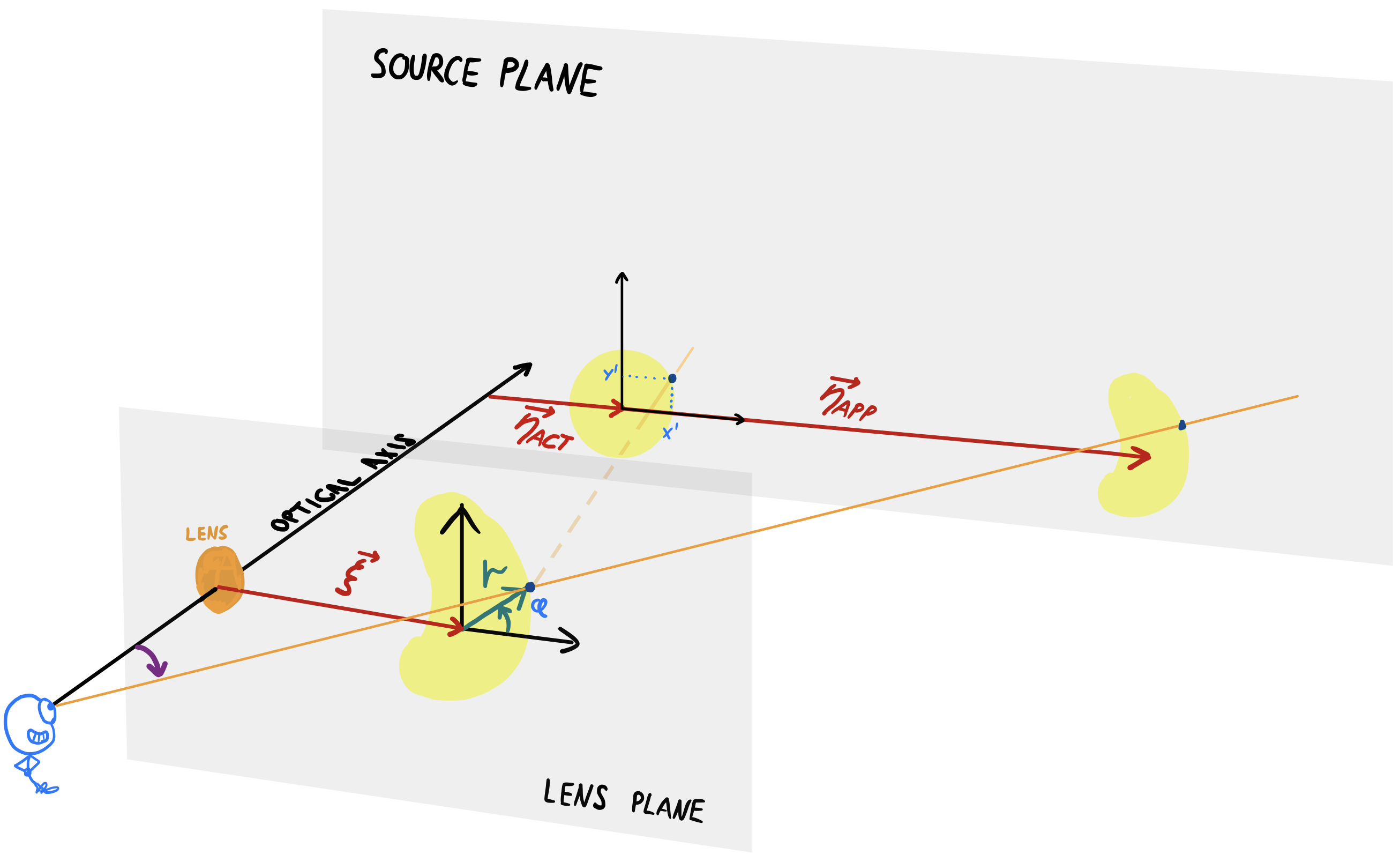}
   \end{center}
    \caption{The figure shows the set-up for the model used. In particular, the local coordinate systems used in the source plane and lens plane are shown. Compare with Figure~\ref{fig:model_2D}.}
    \label{fig1}
\end{figure*}

The Roulette formalism was introduced by 
\citet{clarkson16a}, and
\citet{Clarkson_2016_II} provide complete details. 
Our presentation below will differ a little from conventional
presentations in physics, in order to emphasise computable
functions which can be used to calculate the distorted image.
Readers who want a fuller understanding of the algebraic model
should consult Clarkson's original work.

We study two distant objects in the universe, namely the
(gravitational) lens $L$ at distance $D_\textrm{L}$ from Earth
and the (light) source $S$ at distance $D_\textrm{S}$.
Adopting the \textit{thin-lens approximation}, we assume that
the lens mass is concentrated in a plane orthogonal on the line
of sight through its centre.
The source image is considered only as the 2D projection (image) 
of its emitted light.
With astronomical distances and a relatively small angle of view,
we can assume planar projections; this is known as the \textit{flat-sky
approximation}.
We consider two different images of the source.
The source image is the ideal projection,
as it would have been observed absent any obstructions.
The distorted image is the image as it can be observed 
when light is deflected by the lens.
\begin{table*}
   \begin{minipage}{\textwidth}
\begin{align}
   \alpha_s^m&= - \frac1{2^{\delta_{0s}}} D_\textrm{L}^{m+1}
   \sum_{k=0}^m\binom{m}{k}\left({\mathcal{C}}_s^{m(k)}\partial_{\xi_1}
                                +{\mathcal{C}}_s^{m(k+1)}\partial_{\xi_2}\right)
                      \partial_{\xi_1}^{m-k}\partial_{\xi_2}^k\psi,
                      \label{Alpha}\\
   {\mathcal{C}}_s^{m(k)}&=\frac{1}{\pi}\int_{-\pi}^{\pi}{\rm d}\phi\sin^k\phi\cos^{m-k+1}\phi\cos s\phi\label{C},\\
   \beta_s^m&=-D_\textrm{L}^{m+1}\sum_{k=0}^m\binom{m}{k}\left({\mathcal{S}}_s^{m(k)}\partial_{\xi_1}+{\mathcal{S}}_s^{m(k+1)}\partial_{\xi_2}\right)\partial_{\xi_1}^{m-k}\partial_{\xi_2}^k\psi\label{Beta} \\
   {\mathcal{S}}_s^{m(k)}&=\frac{1}{\pi}\int_{-\pi}^{\pi}{\rm d}\phi\sin^k\phi\cos^{m-k+1}\phi\sin s\phi.\label{S}
\end{align}
   \end{minipage}
   \caption{Constitiuent definitions for the distortion function.}
   \label{tab:eq}
\end{table*}

The observed lensing is decomposed into two steps, as shown in Figure~\ref{fig1}.
The first step is a translation (deflection), corresponding to the difference 
$\boldsymbol{\Delta\eta}$ between actual ($\boldsymbol{\eta}_\textrm{act}$)
and apparent ($\boldsymbol{\eta}_\textrm{app}$) source-plane position. In the roulette formalism, this translational part of the lensing is given as
\begin{align}
\label{Deta}
   \boldsymbol{\Delta\eta}
   & =\boldsymbol{\eta}_\textrm{app}-\boldsymbol{\eta}_\textrm{act}
   =-D_\textrm{S}\cdot(\alpha^0_1,\beta^0_1),
\end{align}
where $(\alpha^0_1,\beta^0_1)$ is a vector of roulette amplitudes, as
defined in Table~\ref{tab:eq}.
The second step is the actual, non-linear distortion.
The distorted image is drawn in a local co-ordinate system in the lens
plane, centred at $\boldsymbol{\xi}=(\xi_1,\xi_2)$, which corresponds to 
$\boldsymbol{\eta}_\textrm{app}$ in the source plane.
We write $\xi=|\boldsymbol{\xi}|$ for the distance between the distorted
image and the lens in the lens plane.
Since $\boldsymbol{\xi}$ and $\boldsymbol{\eta}_{\mathrm{app}}$ lie on the same
line through the viewpoint (cf.\ Figure~\ref{fig:model_2D}), we have
\[ \xi = |\boldsymbol{\xi}| = \frac{D_\textrm{L}}{D_\textrm{S}}\cdot|\boldsymbol{\eta}_{\mathrm{app}}|.\]
Following Clarkson, we use polar co-ordinates $(r,\phi)$ for the
distorted image.
The source image is described in Cartesian co-ordinates $(x^\prime,y^\prime)$ centered
at $\boldsymbol{\eta}_\textrm{act}$ in the source plane.
Thus the light observed at a position (pixel) $(r,\phi)$ is drawn from
a different position (pixel) $(x',y')=\mathcal{D}$$(r,\phi)$ in the source image.
From~Eq.~48 in \citet{Clarkson_2016_II} it is possible to show that 
% \textbf{TODO} CHECK 
the mapping $\mathcal{D}$ is given as
\begin{align}
   % \begin{split}
      \frac{D_{\mathrm{L}}}{D_{\mathrm{S}}}\cdot
   \begin{bmatrix} x' \\ y' \end{bmatrix} &=
   r\cdot\begin{bmatrix} \cos\phi \\ \sin\phi \end{bmatrix} 
    \label{eqn:general mapping}
      + \sum_{m=1}^{\infty} \frac{r^m}{m!\cdot D_{\mathrm{L}}^{m-1}}
      F_s^m 
\end{align}
          {where}
\begin{align}
      F_s^m &=
      \sum_{s=0}^{m+1} c_{m+s}
       \left(\alpha_s^m \boldsymbol{A}_{s} + \beta_s^m \boldsymbol{B}_{s} \right) 
       \begin{bmatrix} C^+ \\ C^- \end{bmatrix}
          \\
   % \end{split}
   C^\pm &= \pm \frac{s}{m+1},\\
   c_{m+s} &= 
      \frac{1 - (-1)^{m+s}}{4} =
   \begin{cases}
      0, \quad m+s \text{ is even},\\
      \frac12, \quad m+s \text{ is odd},
   \end{cases}\\
    \boldsymbol{A}_{s} &= \begin{bmatrix}
    \cos{(s-1)\phi} & \cos{(s+1)\phi} \\ 
    -\sin{(s-1)\phi} &  \sin{(s+1)\phi} \end{bmatrix},
    \\
    \boldsymbol{B}_{s} &=
    \begin{bmatrix} 
        \sin{(s-1)\phi} & \sin{(s+1)\phi} \\
        \cos{(s-1)\phi} & -\cos{(s+1)\phi} 
    \end{bmatrix}.
\end{align}
The coefficients $\alpha_m^s$ and $\beta_m^s$ depend on the lens potential
$\psi(\xi_1,\xi_2)$, from which one may derive the physical properties of the lens.
The general formulae are shown in Table~\ref{tab:eq}.
In practice the sum in \eqref{eqn:general mapping} has to be
truncated by limiting $m\le m_0$ for some $m_0$.

A general implementation for arbitrary $\psi$ would be intractible,
but for many common lens models, it is possible to derive computationally
tractible forms.
The two simplest, but yet very popular, lens models are the 
point mass and singular isothermal sphere (SIS). Confer e.g. with~\cite{bok:schneider92_SEF}, Sections 8.1.2 and 8.1.4 for more on the point-mass and SIS profiles, respectively.
For the point mass, an exact model exists, and we have implemented
both this, and its Roulette approximation.
For SIS, there is no exact model, and we have implemented it in 
the Roulette formalism.

\begin{comment}
In the above, $\partial_{\xi_i}\equiv\frac{\partial}{\partial_{\xi_i}}$. The function $\psi$ is called the lens potential, and contains information about the lens. In particular, the dimensionless surface-mass density may be calculated therof. But also the deflection angle, and higher-order distortions are all derived from the information contained in $\psi$.

Implementing this model in its full generality is intractable at present, and we concentrate here on a couple of lens profiles which allow for extensive simplification. The first of these, is the point-mass lens, which is chiefly interesting because of its analytically tractable nature. As such, it is a useful toy model that provides important grounds for debugging and proof of concept. Its potential is given by
\begin{equation}
   \psi(\xi)=\psi_0\ln(\xi), CHECK \textbf{TODO}
\end{equation}
The second profile is the singular isothermal sphere profile, which is a more realistic model for a galaxy lens. For instance, it is capable of reproducing flat rotation curves, which seems to be a signature of spiral galaxies. Its potential is given by
\begin{align}
   \psi(\xi) &= \frac{R_\textrm{E} \xi}{D_\textrm{L}^2},
   \label{eqn:psi_gamma}
\end{align}
where $R_\textrm{E}$ is the so-called Einstein radius and $\xi=\sqrt{\xi_1^2+\xi_2^2}$ is the distance from the centre of mass of the lens.
\end{comment}

% \section{Lens Models and Gritty Details}
% \label{section:formulas}

\subsection{Point-mass lens}

Without loss of generality, one may assume that the centre of mass of the source is located on the positive $x$-axis.
Using the general equations of \citet{Clarkson_2016_II}, it is
straight forward to find the following formula for point-mass lenses
as a special case:
\begin{align}
   \begin{split}
      \frac{D_\textrm{L}}{D_\textrm{S}} \begin{bmatrix} x' \\ y' \end{bmatrix} &=
       r \begin{bmatrix} \cos{\phi} \\ \sin{\phi} \end{bmatrix} 
          \\&
     - \frac{R_\textrm{E}^2}{\xi} \sum_{m=1}^\infty(-1)^m
     \left(\frac{r}{\xi}\right)^m 
     \begin{bmatrix} \cos(m\phi) \\ -\sin(m\phi) \end{bmatrix}.
   \end{split}
\label{eqn:point sum}
\end{align}
In the above, $R_\textrm{E}$ is the Einstein radius, which is determined
by the gravity (or mass) of the point-mass lens, and thus determines the strength
of the lensing effect.
An approximation of the mapping can be calculated using the sum from $m = 1$
to some finite number $m_0$ with increasing accuracy as $m_0 \rightarrow \infty$.
% This will produce a distorted image with $(m_0+1)$ spurious images and will only
% be valid within $r<\xi$.
This model will be referred to as the 
\textit{finite point-mass model}. Using analytic continuation,
the infinite sum can be calculated and extended outside this region.
Using  geometric series, it can be written in closed form as follows 
\citep{Clarkson_2016_II}:
\begin{align}
   \begin{split}
    \frac{D_\textrm{L}}{D_\textrm{S}} \begin{bmatrix} x' \\ y' \end{bmatrix} 
       &= r \begin{bmatrix} \cos{\phi} \\ \sin{\phi} \end{bmatrix} 
          \\&
       + \frac{R_\textrm{E}^2r}{r^2 + \xi^2 + 2r\xi\cos(\phi)} \begin{bmatrix} 
       \frac{r}{\xi} + \cos(\phi) \\ -\sin(\phi) \end{bmatrix}
   \end{split}.
\label{eqn:point closed}
\end{align}

This is a standard result in the case of a point mass, and is not a result unique to 
the Roulette formalism.
This model will be referred to as the \textit{exact point-mass model}.
The apparent position by the following well-known formula,
\begin{align}
    %% \frac{D_\textrm{S} R_\textrm{E}^2}{D_\textrm{L} \xi}
   |\boldsymbol{\eta}_{\mathrm{app}}|
   =
   \frac{|\boldsymbol{\eta}_{\mathrm{act}}|}{2}
   + \sqrt{ \frac{|\boldsymbol{\eta}_{\mathrm{act}}|^2}{4} + 
            \big(\frac{D_\textrm{S}R_\textrm{E}}{D_\textrm{L}}\big)^2 }.
\label{eqn:shift}
\end{align}

\subsection{General recursive formulae}

A key element of the Roulettes formalism is recursive expressions
for the amplitudes $\alpha_s^m$ and $\beta_s^m$.
Proofs are given by \citet{normann2020}.
The base case is given as,
\begin{align}
    \label{eqn:alpha01}
        \alpha_1^0 = -D_\textrm{L} \frac{\partial \psi}{\partial \xi_1}
    \\
    \label{eqn:beta01}
        \beta_1^0 = -D_\textrm{L} \frac{\partial \psi}{\partial \xi_2}
\end{align}
The recursive relations are given as
\begin{align}
   \alpha_{s+1}^{m+1} &= (C_+^+)_{s+1}^{m+1} (\frac{\partial \alpha_s^m}{\partial \xi_1} 
    - \frac{\partial \beta_s^m}{\partial \xi_2})
    \label{eqn:recursion1}
    \\
     \beta_{s+1}^{m+1} &= (C_+^+)_{s+1}^{m+1} (\frac{\partial \beta_s^m}{\partial \xi_1} + \frac{\partial \alpha_s^m}{\partial \xi_2})
    \label{eqn:recursion2}
    \\
       \alpha_{s-1}^{m+1} &= (C_-^+)_{s-1}^{m+1} (\frac{\partial \alpha_s^m}{\partial \xi_1} + \frac{\partial \beta_s^m}{\partial \xi_2})
    \label{eqn:recursion3}
    \\
     \beta_{s-1}^{m+1} &= (C_-^+)_{s-1}^{m+1} (\frac{\partial \beta_s^m}{\partial \xi_1} - \frac{\partial \alpha_s^m}{\partial \xi_2})
    \label{eqn:recursion4}
\end{align}
with
\begin{align}
	(C_+^+)_s^m &= 2^{\delta_{0(s-1)}} \frac{m + 1}{m + 1 + s} D_\textrm{L}
            \label{eqn:c++}
	    \\
	    (C_-^+)_s^m &= 2^{-\delta_{0s}} \frac{m + 1}{m + 1 - s} D_\textrm{L}
            \label{eqn:c+-}
\end{align}
The astute reader may notice that amplitudes for even sums $s+m$
cannot be found through these relations.
However, the contribution from these terms are equal to zero,
because of the factor $c_{m+s}$ in Equation \eqref{eqn:general mapping}.
In other words, one can calculate all the amplitudes needed from the aforementioned relations.

\subsection{The Singular Isothermal Sphere (SIS)}

The SIS model is somewhat similar to the point-mass model as they both 
have circular symmetry.
The SIS-model however, assumes that the mass of the GL is distributed
in a spherical shape rather than concentrated at a single point.
This means that the final simplifications that was used to get the
simple equations \eqref{eqn:point sum} and \eqref{eqn:point closed}
cannot be used for the SIS model.
However, we can use the recursive formulae from the previous subsection,
with the lens potential given as
%% \footnote{It should be noted here that it is not clear to us how to arrive safely at equation 134 in~\cite{Clarkson_2016_II}, as there seems to be a factor of 1/2 missing. However, the results are consistent with e.g.~\cite{bok:schneider92_SEF}, so we trust it to be correct. Given that it is correct,~\eqref{psiSIS} follows, and $R_{\mathrm{E}}$ corresponds to $\xi_0$ defined in equation~8.34a in~\cite{bok:schneider92_SEF}.}
\begin{align}
\label{psiSIS}
\psi_\textrm{SIS}(\xi)=\frac{R_\textrm{E}}{D_\textrm{L}^2}\xi.
\end{align}
In this case, the Einstein radius $R_\textrm{E}$ depends not only on the total mass,
but also on the size of the SIS lens.  In general it is determined by the
mass distribution of the lens.

\begin{remark}\label{rem1}
   Readers who inspect the source code will note that we use
   have omitted the factors $D_L$ in 
   $\psi$, $\alpha_s^m$ and $\beta_s^m$ ($C_\pm^+$),
   and the right hand side of \eqref{eqn:general mapping}.
   The reason for this is that they all cancel out.
   Verifying this is tedious but straight forward.
\end{remark}

The formula for the apparent position is also different. From Eq.~\eqref{Deta} it follows that
\begin{align}
   |\boldsymbol{\eta}_{\mathrm{app}}| =
   |\boldsymbol{\eta}_{\mathrm{act}}| + \frac{D_\textrm{S}R_\textrm{E}}{D_\textrm{L}},
\label{eqn:shift:sis}
\end{align}
and consequently
\begin{align*}
   \xi = 
   \frac{D_{\mathrm{L}}}{D_{\mathrm{S}}}\cdot|\boldsymbol{\eta}_{\mathrm{app}}| =
   \frac{D_{\mathrm{L}}}{D_{\mathrm{S}}}\cdot|\boldsymbol{\eta}_{\mathrm{act}}|
   + R_{\mathrm{E}}.
\end{align*}